\documentclass{article}

\usepackage{bbm}
\usepackage{bm}
\usepackage{amsfonts}
\usepackage{amssymb}
\usepackage{amsthm}
\usepackage{amsmath}
\usepackage{xcolor} 
\usepackage{hyperref}
\usepackage{float}
\usepackage{tikz-cd}
\usepackage{dsfont}
\usepackage{enumitem}
\usepackage[boxruled, lined, vlined]{algorithm2e}
\usepackage{array}
\usepackage{url}
\usepackage{natbib}
\usepackage{graphicx}
\usepackage{subcaption}
\usepackage{booktabs}
\usepackage{physics}

\usepackage{multirow} 
\usepackage{lineno}
\definecolor{linkcolor}{RGB}{83,83,182}
\hypersetup{
    colorlinks=true,
    citecolor=linkcolor,
    linkcolor=linkcolor
}

\usepackage{arxiv}
\pdfoutput=1 
\allowdisplaybreaks

\title{Discrimination loss vs. SRT: A model-based approach towards harmonizing speech test interpretations}

\author{
  Mareike Buhl$^{1,*}$, Eugen Kludt$^{2,4}$, Lena Schell-Majoor$^{3,4}$, Paul Avan$^{1,\#}$, Marta Campi$^{1,\#}$ \\ \\ 
  $^{1}$Institut Pasteur, Université Paris Cité, Inserm, Institut de l’Audition, IHU reConnect, F-75012, Paris, France \\ \\
  $^{2}$Department of Otorhinolaryngology, Hannover Medical School, Hannover, Germany \\ \\
  $^{3}$Medical Physics, Carl von Ossietzky Universität Oldenburg, Oldenburg, Germany \\ \\
  $^{4}$Cluster of Excellence Hearing4all, Oldenburg \& Hannover, Germany \\ \\
  $^{*}$Corresponding author: Mareike Buhl, mareike.buhl@pasteur.fr \\ \\ 
  $^{\#}$These authors contributed equally.
  }

\begin{document}
\maketitle

\newpage
\begin{abstract} % 200 words
\textbf{Objective:} Speech tests aim to estimate discrimination loss or speech recognition threshold (SRT). This paper investigates the potential to estimate SRTs from clinical data that target at characterizing the discrimination loss. Knowledge about the relationship between the speech test outcome variables--conceptually linked via the psychometric function--is important towards integration of data from different databases.\\ 
\textbf{Design:} Depending on the available data, different SRT estimation procedures were compared and evaluated. A novel, model-based SRT estimation procedure was proposed that deals with incomplete patient data. Interpretations of supra-threshold deficits were assessed for the two \textit{interpretation modes}.\\ 
\textbf{Study sample:} Data for 27009 patients with Freiburg monosyllabic speech test (FMST) and audiogram (AG) results from the same day were included in the retrospective analysis.\\ 
\textbf{Results:} The model-based SRT estimation procedure provided accurate SRTs, but with large deviations in the estimated slope. Supra-threshold hearing loss components differed between the two interpretation modes.\\ 
\textbf{Conclusions:} The model-based procedure can be used for SRT estimation, and its properties relate to data availability for individual patients. All SRT procedures are influenced by the uncertainty of the word recognition scores. In the future, the proposed approach can be used to assess additional differences between speech tests. 

\end{abstract}

%%%%%%%%%%%%%%%%%%%%%%%%%%%
%%%%%%%%%%%%%%%%%%%%%%%%%%%
%%%%%%%%%%%%%%%%%%%%%%%%%%%
\section{Introduction}
%%%%%%%%%%%%%%%%%%%%%%%%%%%
%%%%%%%%%%%%%%%%%%%%%%%%%%%
%%%%%%%%%%%%%%%%%%%%%%%%%%%
During routine audiological patient care, large amounts of patient data are collected and stored in clinical databases, with varying choice of audiological tests or test conditions across different clinic locations. If these data were analyzed together, information about many patients worldwide could be exploited to extract knowledge and to obtain a representative overview of existing patient patterns, and statistical relationships between diagnostic and rehabilitation information could be revealed \citep{saak_flexible_2022}. Such (hearing-impaired) population-representative knowledge holds a large potential for research and advancing hearing healthcare. 
Data in clinical databases represent the current status of how patients are characterized for clinical purposes. In contrast to data collected in research studies, the number of patients is much higher, while data quality may be influenced by time constraints and specific needs of individual patients. 
A common way to interpret and use data from different sources is essential to compare and integrate knowledge, towards enabling big data analysis. 
This paper investigates an audiological research question on routine clinical data. The question of comparing speech test outcomes between the speech recognition threshold ($SRT$) and the maximum word recognition score ($WRS_{max}$) at the same time provides an important step for future integration of data from different audiological databases.

A key application in audiology is the evaluation of hearing device indication criteria and -benefit assessment, which is determined by clinical guidelines (such as \cite{ha_guidelines, ci_guidelines}). Depending on the country, but also on the clinic or audiological center, different speech tests are employed to assess hearing device indication criteria. For clinicians and patients, a standardized interpretation of different speech tests is important to compare data across or within patients, and for providing a fair assessment of hearing device eligibility, which should not depend on the conducted test. 

Mono- or disyllabic word tests in quiet are the speech tests that are most commonly conducted in different countries, such as the German Freiburg monosyllabic speech test (FMST, \cite{hahlbrock1953sprachaudiometrie}) or the French Lafon test \citep{Lafon1}.  
Sentence tests in noise, such as Matrix-type tests \citep{birger_kollmeier_multilingual_2015} or speech tests with meaningful sentences \citep{soli_assessment_2008,kollmeier1997development} are increasingly being used, for example for the assessment of hearing device benefit, and have recently been included in clinical guidelines \citep{ha_guidelines,ci_guidelines,haute_autorite_de_sante_aides_nodate}. Differences between employed speech material, noise condition and -level, or language render comparisons of speech test outcomes difficult. 

Furthermore, a more conceptual and fundamental difference between speech tests is the targeted outcome variable while conducting the test, which relates to different interpretations of the speech test result. These outcome variables are the $SRT$ and the $WRS_{max}$, from which the discrimination loss is derived. In the following, we denote these two variables and related interpretations as \textit{interpretation mode}. 

While sentence tests typically target an $SRT$, word tests are typically aiming at $WRS_{max}$, although both interpretation modes are generally used. For example, German guidelines indicate $WRS$ criteria for the FMST and $SRT$ criteria for the German Matrix test or Göttingen sentence test \citep{ha_guidelines}, while French guidelines relate all speech audiometry results to the $SRT$ (relative to normal-hearing $SRT$) \citep{haute_autorite_de_sante_aides_nodate}.  

Therefore, a first step towards comparing speech test outcomes, and the main objective of this paper, is to understand the relationship between the two interpretation modes, namely $SRT$ and $WRS_{max}$. The analysis of a large clinical database allows to obtain a complete, realistic overview of existing combinations of the two interpretation modes, as well as an idea about the frequency of the different patterns.   
 
Both \textit{interpretation modes} are derived from a patient's psychometric function. This assumed underlying function is characterized by three key parameters: the $SRT$ (the level at which 50\% speech intelligibility is achieved), the slope $s$ at $SRT$ (describing the change of speech intelligibility with level), and the maximum word recognition score $WRS_{max}$. Understanding how these parameters relate to each other is essential before we can meaningfully compare speech test results that use different outcome measures. 

$WRS_{max}$ and $SRT$ both provide an interpretation of a supra-threshold hearing loss component, but the exact relationship or potential correspondence of these interpretations is unclear in literature (e.g., \cite{hoth_freiburger_2016}). A frequently used concept was proposed by Plomp \citep{plomp_auditory_1978}: The Plomp model describes hearing loss as being composed of an audibility ($A$) and a distortion ($D$) component. While the $A$ component describes the part of the hearing loss that can potentially be compensated by amplification, the $D$ component describes an additional supra-threshold component that affects clarity of the audible (amplified) speech, and thus speech intelligibility \citep{plomp_auditory_1978}. 

$WRS_{max}$ describes the maximum speech intelligibility that is achieved by a patient at the corresponding level, representing the performance that should be obtained with hearing devices at a comfortable level, due to amplification. For example, \cite{hoppe_maximum_2019} showed that $WRS_{max}$ predicts the lower limit of aided $WRS$ after cochlear implantation, however not directly predicting the aided $WRS$ as the achieved aided $WRS$ was found to be equal or higher than $WRS_{max}$. To estimate $WRS_{max}$, word recognition scores are assessed at several fixed levels, which are increased until the maximum performance or the audiometer limit is reached. $1-WRS_{max}$ corresponds to the discrimination loss, being related to cochlear damage which leads to a "loss of information-carrying capacity" \citep{halpin_article_2009}. Both the discrimination loss and Plomp's $D$ component represent hearing deficits that cannot be compensated by amplification, suggesting a potential theoretical relationship that remains to be further investigated.   

The $SRT$ characterizes the level or signal-to-noise ratio where 50\% of the presented words or sentences are correctly understood. Together with the slope of the psychometric function, the $SRT$ allows for a level-dependent assessment of speech intelligibility performance at more relevant levels for daily-life communication, as well as to compare the $SRT$ to speech intelligibility expected from the audiogram, thus investigating if a supra-threshold component of hearing loss is present \citep{hoth_freiburger_2016,katz2015handbook}.
Plomp \citep{plomp_auditory_1978} described the influence of $A$ and $D$ on $SRT$ measurements depending on the noise level at which a speech test is conducted. At the lower limit of speech in quiet, the $SRT$ loss (the difference between measured and normal-hearing reference $SRT$ for a given speech test) is given by $A+D$, which means that the two components cannot be disentangled alone with the speech test. When compared with the audiogram, the $A$ component can be estimated and the $D$ component derived. With increasing noise level, the $A$ component is less prominent due to increasing audibility of the noise masking the hearing threshold, and at a certain noise level, the $SRT$ loss is only given by $D$. Hence, in the latter case, if the noise level is high enough relative to the $A$ component of hearing loss, the speech test purely characterizes the supra-threshold hearing loss component \citep{plomp_auditory_1978}. Figure \ref{fig:data_flowchart} (B) illustrates the relationship of $A$ and $D$ with the $SRT$.

Since the underlying relationship between the two interpretation modes is given by the psychometric function, a transformation between the two outcome metrics should be possible if a speech test is conducted at sufficient and suitable levels to characterize all psychometric function parameters for individual patients. Data collection for (adaptive) $SRT$ estimation rarely provides data points at levels at which $WRS_{max}$ is expected, while clinical data collection aiming at $WRS_{max}$ estimation can include lower speech levels close to the $SRT$. For this reason, we chose to analyze a clinical database with data collection in the $WRS_{max}$ interpretation mode, employing the Freiburg monosyllabic speech test (FMST; \cite{hahlbrock1953sprachaudiometrie}). 

The FMST is the most-used speech intelligibility test in Germany \citep{hoth_freiburger_2016}, as well as for the German-speaking population in Switzerland \citep{kompis_uberprufung_2006}. The FMST has been extensively characterized in the literature, as summarized by \cite{hoth_freiburger_2016}. \cite{baljic2016evaluation} showed that the different test lists do not provide equivalent speech intelligibility when presented at the same level, and proposed to exclude four test lists from clinical use. \cite{holube_modelling_nodate} modeled the test-retest reliability of the FMST, they found that the confidence interval for the true $WRS$ value given the obtained measurement result depends on the obtained $WRS$, with largest confidence intervals of up to $\pm$ 17.1\% at 50\% speech intelligibility and smallest confidence intervals at 0 and 100\%. Despite such known disadvantages, the FMST is still commonly used and relevant in clinical practice. First, for many patients, especially with more severe hearing loss, an estimate of the $WRS_{max}$ can be more important than a precise characterization at different levels. Second, alternative speech tests have different properties such as using sentences instead of words, or are proposed to be conducted in noise. These properties make them more suitable for other groups of the audiological patient population, with different hearing loss characteristics such as supra-threshold deficits or additional demands on, e.g., cognitive factors. As a consequence, comparison between different speech tests is difficult, and not one test may be optimal for all patients \citep{hoth_freiburger_2016}. 

For the FMST conducted in quiet at fixed speech levels per test list, data availability for different speech levels is determined by two factors: the chosen presentation levels during measurement and the $WRS$ result obtained at the respective previous level. In the ideal case, a psychometric function fit could provide an $SRT$ estimate if three levels (in the appropriate ranges of the function) were measured, with $SRT$, $s$, and $WRS_{max}$ as free parameters. However, in the $WRS_{max}$ interpretation mode often only one or two test lists at different levels are performed, providing insufficient data points to characterize both $SRT$ and slope parameters. In this case, a \textit{model-based} approach could provide additional knowledge to complement the measured data. 

By design, $WRS_{max}$ is always measured, providing one data point that is consistently available. However, without additional data points at lower levels, we cannot estimate $SRT$ and slope solely based on the measured data. To estimate both parameters, two additional data points are needed in a level range that leads to $WRS$ results in the assumed linear range of the psychometric function, at audible levels below the level corresponding to $WRS_{max}$. This is denoted as \textit{slope area} in the following. This $SRT$ estimation procedure is called the \textit{empirical slope SRT estimation}, and the corresponding patients are denoted as \textit{fully-determined slope} patients.     

When only one additional data point is available beyond $WRS_{max}$, we propose estimating the $SRT$ using a linear fit with a slope derived from the speech intelligibility index (SII). We call this model-based approach the \textit{SII-slope-based} $SRT$ estimation and classify these cases as \textit{half-determined slope} patients. 

The SII is a speech intelligibility model that estimates audible cues of a speech signal presented at a certain level, with weighting applied across different frequency bands (band importance function) \citep{ansi_1997}. It can be individualized by incorporating the patient's audiogram - higher hearing thresholds result in lower SII values. The SII accounts for specific speech test characteristics through the spectrum of the speech material used. Since the SII links audiogram with speech intelligibility and varies with presentation level, it provides the potential to estimate the psychometric function slope.
 
Varying the speech level produces an SII-level curve that behaves similarly to a psychometric function, showing a linear slope (in SII per dB) within a certain level range. The SII slope is assumed to be proportional to the psychometric function's slope because both SII and WRS represent speech intelligibility as a function of speech level. This relationship aligns with findings from \cite{smits_interpretation_2011}, who validated this proportionality for speech tests in stationary noise due to the linear nature of both slopes. A transfer function between WRS and SII can be established when a reference condition is known - specifically, when a measured $SRT$ is available that corresponds to the respective SII calculation condition \citep{ansi_1997}.

For our proposed $SRT$ estimation procedure, we only use the slope, eliminating the need for absolute SII predictions. While the validity of the SII-based slope for individual listeners is constrained by the SII's inherent properties and assumptions (cf. Section 2.3), we use the SII because it provides a consistent way to relate speech intelligibility and audibility across patients.

In summary, the current paper aims to investigate the potential to estimate an $SRT$ from measured data, as well as by employing our proposed model-driven approach. For this purpose, the analysis was performed on a clinical database targeting the maximum intelligibility (containing Freiburg monosyllabic speech test in quiet and pure-tone audiogram). Moreover, the relationship between the two \textit{interpretation modes} was investigated. More specifically, the following research questions are addressed: 
\begin{itemize}
    \item[\textbf{RQ 1}] For which patients can we derive an SRT estimate based only on empirical data (\textit{fully-determined slope} patients, \textit{empirical slope} procedure), and with which error? This is considered as ground truth in comparisons. 
    \item[\textbf{RQ 2}] For patients with incomplete data (not suitable for $SRT$ estimation only based on empirical data), can we estimate a plausible $SRT$ based on model-driven assumptions (\textit{half-determined slope} patients, \textit{SII-slope-based} procedure)? With which error? For which patients is no $SRT$ estimation feasible?  
    \item[\textbf{RQ 3}] If a difference between the two estimated SRTs exists: Are the parameters used for the SII-slope-based procedure able to explain the $SRT$ difference? The rationale is to understand which factors influenced the $SRT$ estimation in the proposed model-driven procedure.  
    \item[\textbf{RQ 4}] Do we find a relationship between supra-threshold hearing loss components characterized by the two \textit{interpretation modes} for the Freiburg monosyllabic speech test?  
\end{itemize}

%%%%%%%%%%%%%%%%%%%%%%%%%%%
%%%%%%%%%%%%%%%%%%%%%%%%%%%
%%%%%%%%%%%%%%%%%%%%%%%%%%%
\section{Materials and Methods}
%%%%%%%%%%%%%%%%%%%%%%%%%%%
%%%%%%%%%%%%%%%%%%%%%%%%%%%
%%%%%%%%%%%%%%%%%%%%%%%%%%%

\subsection{Database}
\subsubsection{Overview}
The clinical database of the ENT Department at Hannover Medical School (MHH, Germany), a tertiary care university clinic, was utilized for the present analysis. This database contains records of patients who visited MHH since the introduction of the first network-connected digital audiometers in 2002. Data from patients who provided consent for the use of their data in retrospective research were transferred to a scientific database. In this database, patients were pseudonymized and various within-hospital data sources were integrated to link demographic and anamnestic information with audiological data, imaging results, and hearing device-specific details, such as cochlear implant fitting protocols.

\subsubsection{Remote analysis}
The research database can be accessed by members and collaborators of the Cluster of Excellence “Hearing4all” through remote query procedures. A data dictionary and synthetic example data were provided to an external collaborator (MB) to facilitate the development of queries and analysis scripts. These scripts were subsequently reviewed and refined collaboratively with a clinic-based researcher (EK). Both parties ensured that the scripts fully anonymized all outputs. Once this was confirmed, the analysis scripts were executed within the MHH environment on the real dataset. The resulting anonymized data was then securely transferred to the external collaborator for further evaluation. This procedure ensured that no re-identification of patients outside the MHH was possible.

\subsubsection{Data choice for the present study}
For the purpose of this study, a data set was extracted containing pure-tone audiogram and Freiburg monosyllabic speech test (FMST; \cite{hahlbrock1953sprachaudiometrie}) results for in total 32218 patients, conducted on the same day between 2002 and 2023. When measurements at multiple time points were available, the earliest measurement was chosen for the analysis. In addition, information about gender, the date when the testing was performed, as well as age at test date were available. Since the aim was to investigate the relationship between $WRS_{max}$ and $SRT$ within patients, a rather methodological question, all patients were used for analysis (for more details see Section \ref{sec:preproc}).
The pure-tone audiogram (AG) was conducted at the frequencies 250, 500, 1000, 1500, 2000, 3000, 4000, 6000, and 8000 Hz. Digital audiometers (AD2117, Audio-DATA, Germany) equipped with HDA 200 headphones (Sennheiser, Germany) were used to perform audiometric measurements in a soundproof booth. Both audiogram and FMST were conducted monaurally for both ears, with masking applied to the contralateral ear if necessary. Lists of 20 monosyllabic words were presented in quiet, at fixed speech levels of 60, 80, 100, and 110 dB SPL, respectively. For each patient, only those levels were measured that were required to characterize the maximum word recognition score ($WRS_{max}$), therefore the choice of levels and corresponding measured $WRS$ values varies across patients. 
The normal-hearing psychometric function of the FMST in quiet yields an $SRT_{NH}$ of 29.3 dB SPL and a slope $s_{WRS,NH}$ of 4.5\%/dB \citep{baljic2016evaluation}. 

\begin{figure}%[H]
\centering
 \includegraphics[width=1.0\linewidth]{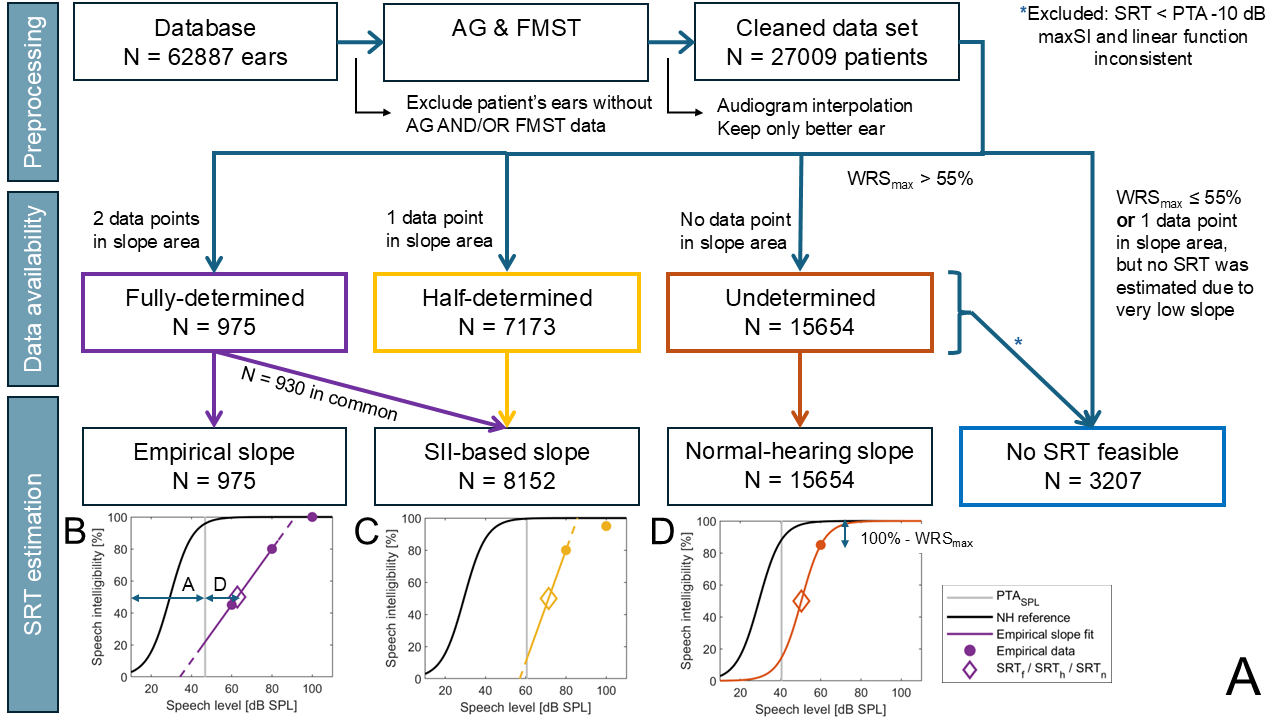}
  \caption{(A) Data flowchart, overview about data cleaning and data availability for different $SRT$ estimation methods. Note that, after excluding $WRS_{max}$, not all remaining data points necessarily fall within the defined \textit{slope area}, meaning the actual number of data points in the slope area can be less than $N_{DP}-1$. The colored boxes represent the total number of patients included in the final analysis, as shown in Figure \ref{fig:maxSI_pta}. Examples of data availability and resulting fit are depicted in (B) for the empirical slope-based, (C) for the SII-slope-based, and (D) for the normal-hearing slope-based $SRT$ estimation procedure, respectively. Continuous lines in subfigures (B) and (C) indicate the slope area. For illustrative purposes, Plomp's $A$ and $D$ component are indicated in subfigure (B), and the discrimination loss $100\%-WRS_{max}$ is visualized in subfigure (D).  Abbreviations: Audiogram (AG), Freiburg monosyllabic speech test (FMST), speech recognition threshold (SRT), pure-tone average (PTA), maximum word recognition score ($WRS_{max}$), speech intelligibility index (SII). }

\label{fig:data_flowchart}
\end{figure}

\subsection{Data preprocessing}

\label{sec:preproc}
All analyses were performed in Matlab 2022b (The Mathworks, including the Curve Fitting Toolbox). The code can be found at \url{https://doi.org/10.5281/zenodo.14634515}. To prepare the data for the remote analysis, several preprocessing steps were performed. Figure \ref{fig:data_flowchart} (A, upper part) visualizes these steps. 
First, the number of available data points for audiogram (frequencies) and FMST (measured speech levels) were calculated to check for completeness, and all patients without any data point in the audiogram or the FMST were excluded. 
Next, missing values in the audiogram were imputed (interpolation based on nearest values of neighboring frequencies, and limiting the allowed range between $-10$ and $120$ dB HL to capture plausible values) as a prerequisite for the following SII calculations. 
The pure-tone average ($PTA$) was calculated based on the frequencies 500, 1000, 2000, and 4000 Hz, and transformed to dB SPL to later allow comparison to the speech test levels. Based on the $PTA$, the better ear was estimated, and only data for the respective better ear of each patient was kept. This resulted in $N = 27009$ patients as starting point for the analysis. 
Histograms for the variables ear, gender, age at measurement, and test date were calculated to extract descriptive information about the data set from the remote analysis. 
Finally, the table containing all patients was additionally filtered for unique audiograms, to reduce computing time in the SII analysis (cf. Section \ref{sec:sii}). The estimated SII slopes were mapped back to all patients in the $SRT$ estimation (cf. Section \ref{sec:srt_estimation}).

\subsection{Categorization of patients for analysis}
\label{sec:pat_cat}
Due to the limited availability of data points in the \textit{slope area}, a complete psychometric function fit, using all three parameters, was only possible for a small subset of patients. Therefore, we analyzed the $WRS_{max}$ and the data in the slope area separately. $WRS_{max}$ was characterized by the empirical $WRS_{max}$ (rightmost data point in Figure \ref{fig:data_flowchart} (B), (C), and (D)). The choice of $SRT$ estimation procedure depended on the number of measured levels and their corresponding $WRS$ values in the \textit{slope area} (cf. Figure \ref{fig:data_flowchart}). The following cases were distinguished:  

\begin{itemize}  
    \item \textbf{Fully-determined slope}. Two data points are available in the \textit{slope area}, corresponding to two measured levels for which a speech intelligibility score in [0.15 0.85] $\cdot WRS_{max}$ was obtained. These points, assumed to be in the linear range of the psychometric function, allowed for $SRT$ estimation based on measured data (cf. Section \ref{sec:emp_procedure}). 
    
    \item \textbf{Half-determined slope}. One data point is available in the \textit{slope area}. While insufficient for empirical slope estimation, additional, model-based knowledge can be used to estimate a best-possible slope (cf. Section \ref{sec:sii_procedure}). 
    \item \textbf{Undetermined slope}. No data point in the slope area. Only $WRS_{max}$ and its corresponding level are available, providing no information for slope estimation (cf. Section \ref{sec:nh_procedure}). 
\end{itemize}

\subsection{SII-based slope calculation}
\label{sec:sii}

For all patients with fully- or half-determined empirical slope (one or more data points in slope area, cf. Figure \ref{fig:data_flowchart}), SII calculations were performed to derive a model-based slope estimate from the SII-level-curve. We assume the SII slope to be proportional to the psychometric function's slope since both SII and $WRS$ represent speech intelligibility as a function of speech level. The SII increases with speech level as more frequency bands become audible, providing additional audible speech cues. Due to this relationship with audibility, the slope depends on two factors: the frequency distribution of the audiogram and the band importance function. 
Flat audiograms yield the highest slope, while differences across frequencies lead to a lower slope, showing a cumulative behavior of the SII curve with increasing level due to audibility. 

For SII calculations, we used test-specific noise generated by superimposing Freiburg monosyllabic test lists \citep{zinner2021vergleich} as speech material. This approach is equivalent to using the speech material of the words, as SII calculations are based on the speech spectrum. To model the quiet condition of the speech test, we set the noise level to -50 dB SPL. The SII was calculated in 21 critical frequency bands and weighted by the speech-in-noise (SPIN) band importance function for comparability with calculations in a related study. The level distortion factor of the SII was enabled to obtain plausible SII upper limits for the slope calculations \citep{ansi_1997}. 
We incorporated each patient's audiogram into the SII calculation to account for individual hearing thresholds and to limit audible cues. To optimize computing time, calculations were performed only on unique audiograms (cf. Section \ref{sec:preproc}).     

To calculate the SII slope, SII calculations were performed at different speech levels. Here, computing time was optimized by choosing relevant levels: For every individual audiogram, the SII was first calculated at four levels to estimate the general dependence on level ($L=10,~40,~70,~100~\text{dB SPL}$). Based on the obtained SII values, the level range was estimated where the linear, positive slope was expected, and the SII was estimated at additional four equidistant levels in this range. We assessed linearity by calculating Pearson's correlation coefficient for each set of three consecutive level-SII-points. The rationale behind this is that a correlation close to 1 is only obtained if the three points are in the linear range of the curve. Additional levels were iteratively added if $R^2 < 0.99$, and the linearity check was repeated. The procedure was stopped if $R^2 \geq 0.99$, or interrupted if the algorithm proposed no additional levels. 

The method uses two equations to estimate the psychometric function slope $s_{WRS}$, the slope for the SII-based $SRT$ estimation procedure. First, the SII slope $s_{SII}$ was calculated using the upper and lower data points ($L_u$,$SII_u$) and ($L_l$,$SII_l$) from the three consecutive data points showing highest correlation (Equation \ref{eq:slope_SII}). Then, to convert the SII slope to the slope $s_{WRS}$ in the $WRS$ domain, a proportionality factor established from the normal-hearing reference slope ($s_{WRS,NH} = 4.5~\%/\text{dB}$ for the Freiburg monosyllabic speech test in quiet; \cite{baljic2016evaluation}) and the SII slope calculated for a zero-threshold audiogram ($s_{SII,NH} = 0.0307~1/\text{dB}$) were used (Equation \ref{eq:slope_SI}).   

\begin{equation}
    s_{SII} = \frac{SII_u-SII_l}{L_u-L_l}
    \label{eq:slope_SII}
\end{equation}

\begin{equation}
    s_{WRS} = s_{SII} \cdot \frac{s_{WRS,NH}}{s_{SII,NH}}
    \label{eq:slope_SI}
\end{equation}

\subsection{SRT estimation}
\label{sec:srt_estimation}

If sufficient data is available, the typical way of estimating an $SRT$ is to fit a psychometric function to the speech audiometry data. Equation \ref{eq:psyfun} describes the relationship between speech intelligibility $WRS$ and level $L$ as a logistic function with the parameters speech recognition threshold ($SRT$) indicating the level required to obtain 50\% speech intelligibility, slope of the psychometric function at the $SRT$ ($s_{WRS}$), and the maximum speech intelligibility ($WRS_{max}$). To fit a function with the three parameters, at least three data points in the respective corresponding level range need to be available. 

\begin{equation}
    WRS(L) = WRS_{max} \frac{1}{1+e^{4 \cdot s_{WRS,NH} \cdot (SRT_{NH}-L) }} 
    \label{eq:psyfun}
\end{equation}

The different $SRT$ estimation approaches depending on the data availability in the \textit{slope area}, as introduced in Section \ref{sec:pat_cat}, are described in detail in the following. In all cases, $SRT$ and $PTA$ were checked for consistency after $SRT$ estimation, and SRTs with $SRT < PTA_{SPL} - 10~\text{dB}$ were excluded. All quantities are distinguished by index $f$ for the empirical slope-based $SRT$ estimation, $h$ for the SII-based-slope $SRT$ estimation, and $n$ for the normal-hearing slope-based $SRT$ estimation.      
\subsubsection{Empirical slope-based SRT estimation}
\label{sec:emp_procedure}
For \textit{fully-determined slope} patients, a linear fit was performed based on the two available data points in the slope area. The $SRT_f$ was obtained by interpolating the resulting linear function, characterized by intercept and empirical slope $s_f$, to $WRS=50\%$. 
    
\subsubsection{SII-based-slope SRT estimation}
\label{sec:sii_procedure}
For \textit{half-determined slope} patients with one available data point in the slope area, a linear fit with fixed slope was performed through this single data point. The slope $s_h$ was determined from the SII slope as given by Equation \ref{eq:slope_SI}, using the individual audiogram in the SII calculations. The $SRT_h$ was obtained by interpolating the resulting linear function (characterized by intercept and SII-based slope $s_h$) to $WRS=50\%$. 
For \textit{fully-determined slope} patients, this $SRT$ estimation was also performed to obtain a comparison for the same patients. In this case, one of the two available data points was chosen, which was audible (defined by the criterion $L>=PTA_{SPL}-10~\text{dB}$ to avoid loss of data points that are still in the range of potential errors), and closer to $WRS=50\%$.

\subsubsection{Normal-hearing slope-based SRT estimation}
\label{sec:nh_procedure}
For \textit{undetermined slope patients}, for which no data point was available in the slope area, and in total only the $WRS_{max}$ data point was available, a psychometric function fit according to Equation \ref{eq:psyfun} with fixed parameters $s=s_{WRS,NH}$ and $WRS_{max} = 100 \%$ was performed. Hence, only the $SRT$ parameter was fitted, with otherwise most simplified assumptions of normal-hearing parameters. Even though this is not expected to yield a precise $SRT$ estimate since the individual slope may differ, it is the only possible procedure given the available data. This $SRT$ serves as comparison, and as an upper limit of a realistic range of SRTs, since any fit with a lower slope would result in a smaller $SRT$. The lower limit is given by the maximum of the normal-hearing reference $SRT_{NH}$ and $PTA_{SPL}-10~\text{dB}$ to incorporate audibility.

\subsubsection{Error calculations} 
\label{sec:errors}

The $SRT$ estimation error ($\Delta SRT$) depends on multiple components in our $SRT$ procedures. For both the empirical and SII-based slope methods, we estimated SRTs usings available data points and slope (Equation \ref{eq:srt}). The level error is assumed as $\Delta L = 0$ since we don't control for its influence on the speech intelligibility measurement, and it is constant. The $SRT$ estimation error $\Delta SRT$ depends on the measured word recognition score $WRS_i$, the $WRS$ error $\Delta WRS_i$ of data point $i$, as well as the slope $s$ and the slope error $\Delta s$ (Equation \ref{eq:delta_srt}).

\begin{equation}
    SRT = L_{WRS,i} - (L_{WRS,i}-L_{SRT}) = L_{WRS,i} - \frac{WRS_i-50\%}{s}
    \label{eq:srt}
\end{equation}
% SRT_h, s_h

\begin{equation}
    \Delta SRT = \frac{1}{s} \cdot \Delta WRS_i + WRS_i - 50\% \cdot \frac{1}{s^2} \cdot \Delta s
    \label{eq:delta_srt}
\end{equation}

The $WRS$ estimation error ($\Delta WRS_i$) varies with the measured speech intelligibility \citep{holube_modelling_nodate}. By modelling the Freiburg monosyllabic speech test in quiet as a Bernoulli experiment, \cite{holube_modelling_nodate} established the 95\% confidence intervals for true value of $WRS$ given the measured $WRS$. The upper and lower limit confidence intervals given in Table 5 of \cite{holube_modelling_nodate} for one test list were used as upper and lower limits in the $\Delta WRS_i$ error calculation, corresponding to the respective achieved $WRS$ at the data point $WRS_i$ for each patient.    

The slope estimation error ($\Delta s$) calculation differs between the two $SRT$ estimation procedures that involve a slope. For \textit{fully-determined slope} estimation, we calculated the slope $s_f$ using upper ($u$) and lower ($l$) data points according to Equation \ref{eq:slope_f}. The corresponding error ($\Delta s_f$) is given by Equation \ref{eq:delta_slope_f}, again with $\Delta L = 0$. 

For \textit{half-determined slope} estimation, the slope $s_h$ was calculated from the SII slope as described in Section \ref{sec:sii}. We estimated the measurement uncertainty by calculating the standard deviation of repeated SII calculations using different random speech signal excerpts. This measurement uncertainty can be influenced by underlying estimation errors, for example of the included audiograms, however, such factors are already incorporated in the calculated measurement uncertainty. To estimate the magnitude and to verify that this error is stable across levels and audiograms, we repeated the SII calculation for all levels and all Bisgaard standard audiograms \citep{bisgaard2010standard} ten times. We used the maximum error $\Delta SII = 0.00084$ for the calculation of $\Delta s_h$, as $\Delta SII$ was very small compared to the modelled SII values. We calculated the resulting slope error ($\Delta s_h$) using Equation \ref{eq:delta_slope_h}, which incorporates the SII uncertainty from both level-SII points into the slope calculation. 

\begin{equation}
    s_{f} = \frac{WRS_u - WRS_l}{L_u - L_l} \\
    \label{eq:slope_f}
\end{equation}

\begin{align}
    \Delta s_{f} &= \left| \frac{\partial s_f}{\partial WRS_u} \right| \Delta WRS_u  + \left| \frac{\partial s_f}{\partial WRS_l} \right| \Delta WRS_l \nonumber \\
                 &= \frac{1}{L_u - L_l} (\Delta WRS_u + \Delta WRS_l )
                 \label{eq:delta_slope_f}
\end{align}

%\begin{align}
%    \Delta s_{f} &= \bigg\rvert \frac{\partial s_f}{\partial WRS_u} \bigg\rvert \Delta WRS_u  + \bigg\rvert \frac{\partial s_f}{\partial WRS_l} \bigg\rvert \Delta WRS_l \nonumber\\
%                 &= \abs{\frac{1}{L_u - L_l} (\Delta WRS_u + \Delta WRS_l)}
%                 \label{eq:delta_slope_f}
%\end{align}

\begin{equation}
    \Delta s_h = \sqrt{2 \cdot \Delta SII^2}
    \label{eq:delta_slope_h}
\end{equation}

For \textit{undetermined slope} patients, we establish $SRT$ error limits differently. The $SRT_n$ resulting from the fit is only influenced by the error $\Delta WRS_i$ of the data point used for the fit. However, the assumption of a normal-hearing slope employed in this procedure, which is the only possible way to derive an $SRT$ in this case, corresponds to an assumed upper limit of the SRT. Hearing-impaired slopes would be lower than the normal-hearing slope, resulting in lower SRTs than the estimated $SRT_n$. Therefore, we define the $SRT$ error as the maximum range of possible SRTs in the context of this paper. The minimum $SRT$ ($SRT_{n,min}$) is assumed as the maximum of $SRT_{NH}$ and $PTA-10~\text{dB}$ according to Equation \ref{eq:srt_n_min}. The maximum range of SRT, considered as the error in this estimation procedure, is then given by Equation \ref{eq:delta_srt_n}. 

\begin{equation}
    SRT_{n,min} = max(SRT_{NH},PTA-10~\text{dB})
    \label{eq:srt_n_min}
\end{equation}

\begin{equation}
    \Delta SRT_{n} = SRT_{n}-SRT_{n,min}
    \label{eq:delta_srt_n}
\end{equation}

\subsection{Plomp evaluation} 
\label{sec:plomp}
We estimated Plomp's $A$ and $D$ components \citep{plomp_auditory_1978} to evaluate the contributions of audibility and supra-threshold hearing deficits captured in our $SRT$ estimation procedures, and to compare these with the discrimination loss ($100 \% - WRS_{max}$). 

The $A$ component corresponds to the $PTA_{SPL}$ (Equation \ref{eq:plomp_A}). The $D$ component is calculated as the difference between $SRT$ and the $A$ component according to Equation \ref{eq:plomp_D}, for all three estimation procedures leading to $SRT_f$, $SRT_h$, and $SRT_n$. Note that $SRT_{NH}$ was always subtracted to represent $SRT$ loss. The error of the $D$ component $\Delta D$ is given by Equation \ref{eq:plomp_dD}, with $\Delta PTA_{SPL} = 5~\text{dB}$. % = \Delta AG 

\begin{equation}
    A = max(PTA_{SPL} - SRT_{NH},0~\text{dB SPL})
    \label{eq:plomp_A}
\end{equation}

\begin{equation}
    D = SRT - SRT_{NH} - A = SRT - PTA_{SPL}
    \label{eq:plomp_D}
\end{equation}

\begin{equation}
    \Delta D = \Delta SRT - \Delta PTA_{SPL}
    \label{eq:plomp_dD}
\end{equation}

\subsection{Statistical analysis}
\subsubsection{Comparison of measured data between \textit{fully-} and \textit{half-determined slope} patients}
 
To validate whether the SII-based procedure (used for \textit{half-determined slope} patients, while direct comparison was only possible for \textit{fully-determined} patients) generalizes appropriately, we compared $PTA$ and $WRS_{max}$ distributions between \textit{fully-} and \textit{half-determined} patient groups using one metric and two statistical tests. The overlapping index \citep{pastore_measuring_2019}, our distribution metric, quantifies the overlap between two normalized probability density functions or histograms in a distribution-free way, yielding values between 0 and 1. For statistical comparison, we employed the Welch test \citep{field2009discovering}, a two-tailed t-test that assesses mean differences while accommodating unequal sample sizes and variances, and the Kolmogorov-Smirnov test \citep{field2009discovering}, which compares the growth rates of empirical cumulative distribution functions. We set the significance level at $\alpha = 0.05$ for both statistical tests. This combination of a metric and statistical tests provides a comprehensive comparison of $PTA$ and $WRS_{max}$ distributions between patient groups, with the statistical tests complementing the quantitative overlap assessment.

\subsubsection{Prediction of SRT difference between empirical slope and SII-based-slope procedures}

To assess if the obtained differences between $SRT_f$ and $SRT_h$ (for \textit{fully-determined slope} patients) can be used to predict a plausible $SRT_h$ for all \textit{half-determined slope} patients, a generalized linear model (GLM, \cite{jiang_linear_2021}) was built.
Our rationale was twofold: first, if successful, this prediction could serve as a correction to improve the SII-slope-based $SRT$ estimation. Second, it helps to identify which variables influence the estimation's prediction error. A successful model could then be applied to correct $SRT$ estimates for all half-determined slope patients through out-of-sample prediction. 

The GLM is defined by Equation \ref{eq:srt_diff_pred}. The SII-based slope $s_h$ and the $WRS$ difference between the employed data point and the estimated $SRT$ ($WRS_i-50\%$) were used as predictor variables since these parameters were used for the SII-slope-based $SRT$ estimation (cf. Equation \ref{eq:srt}). The GLM model was estimated and evaluated in a 10-fold cross-validation. In 10 folds, a model was estimated based on 90\% of the patient data and used to predict the respective hold-out 10\%. The cross-validation error was calculated as RMS value of the mean squared error (MSE) across folds. To evaluate the prediction of $SRT$ differences compared to observed $SRT$ differences, Pearson's correlation coefficient, RMSE, and bias were calculated. 

\begin{align}
    SRT_{diff}  &= SRT_f - SRT_h \nonumber\\
            SRT_{diff}    &\sim \beta_0 + \beta_1 \cdot s_h + \beta_2 \cdot (WRS_i - 50\%)
    \label{eq:srt_diff_pred}
\end{align}

%%%%%%%%%%%%%%%%%%%%%%%%%%%
%%%%%%%%%%%%%%%%%%%%%%%%%%%
%%%%%%%%%%%%%%%%%%%%%%%%%%%
\section{Results}

\subsection{Clinical data availability and clinical interpretation}

\begin{figure}
\centering
 \includegraphics[width=1\linewidth]{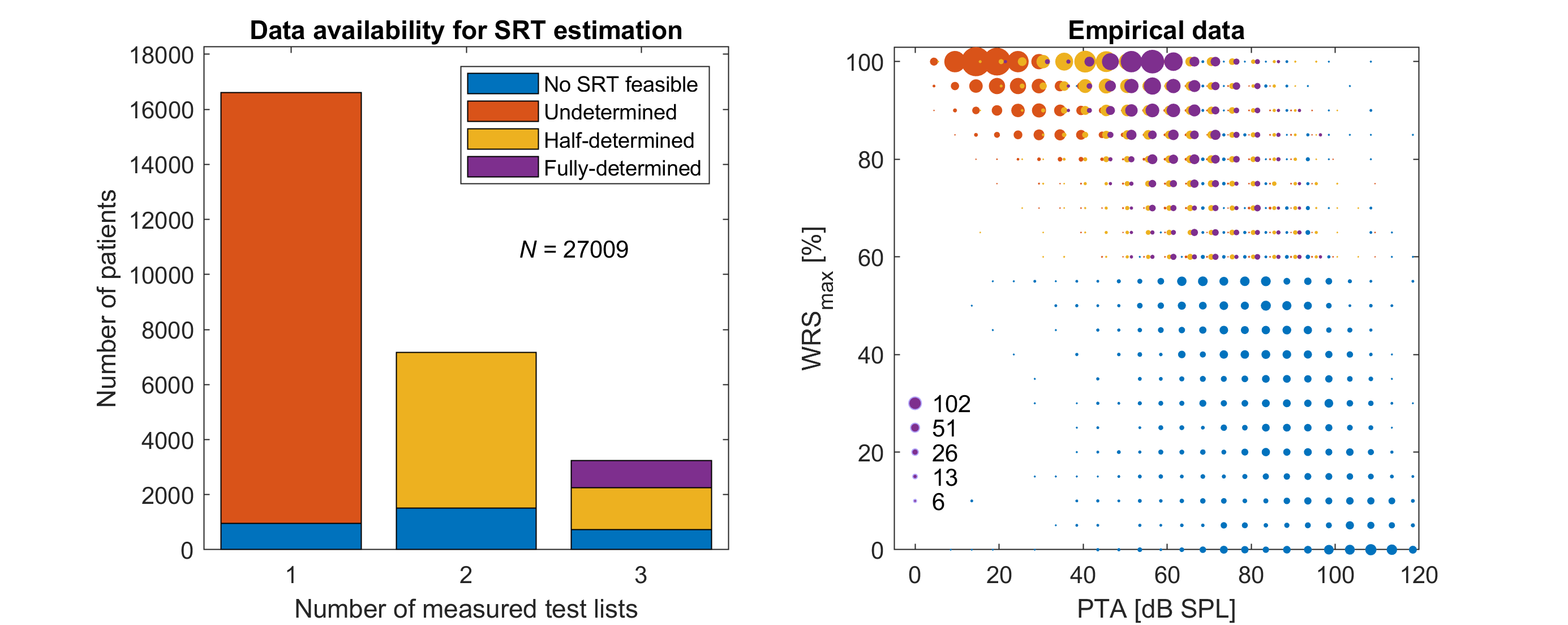}
  \caption{Clinical data availability in the $WRS_{max}$ \textit{interpretation mode}. (A) Number of patients for different numbers of measured FMST test lists. The data availability according to Figure \ref{fig:data_flowchart} is color-coded. (B) Scatterplot of existing combinations of $WRS_{max}$ and $PTA$ in the data set. Colors correspond to the same groups as in (A), the marker size indicates the number of patients in each combination, in logarithmic scaling.} 
\label{fig:maxSI_pta}
\end{figure}

Prior to estimating SRTs, the starting point of available data in the $WRS_{max}$ \textit{interpretation mode} needs to be characterized. Figure \ref{fig:maxSI_pta} (A) displays the number of measured test lists for the $N=27009$ patients who completed both the pure-tone audiogram and the FMST, and are used for analysis after data cleaning and preprocessing (cf. Figure \ref{fig:data_flowchart} for details). The analyzed patient group was balanced in gender ($46\%$ female, $54\%$ male) and better ear ($48\%$ left, $52\%$ right), and had a median age at measurement of 56 years ($IQR = [41~68]~\text{years}$). Figure \ref{fig:maxSI_pta} (B) shows the existing combinations between the clinical outcome measure $WRS_{max}$ and the $PTA$. 
For a small part of these patients, no $SRT$ estimation was feasible since $WRS_{max}$ was too low (blue). $SRT$ estimation with the empirical slope procedure was only possible for 3.4\% of the patients who exhibit two data points in the \textit{slope area}, and three data points in total (fully-determined, violet). About a quarter of the patients have one data point available for $SRT$ estimation with the SII-slope based procedure, with either two or three test lists measured in total (half-determined, yellow). These two groups show generally overlapping ranges in Figure \ref{fig:maxSI_pta} (B), with a smaller $PTA$ range visible for fully-determined patients. For more than half of the patients, only one test list was conducted, in most cases resulting in a $WRS_{max} \geq 80\%$ (undetermined slope, red). These cases do not fulfill, if conducted at the lowest level, the indication criteria for a hearing aid in terms of unaided speech intelligibility, and were therefore not further characterized in the clinical data collection \citep{ha_guidelines}. The patients for which no $SRT$ can be estimated at all (blue), could be CI candidates according to \cite{ci_guidelines} (note that the other parts of the criteria would need to be checked as well, which is beyond the scope of this paper), while the fully- and half-determined patients should further be investigated towards eligibility for hearing aids. 

\subsection{SRT estimation} 

\begin{figure}%[H]
\centering
\includegraphics[width=1.1\linewidth]{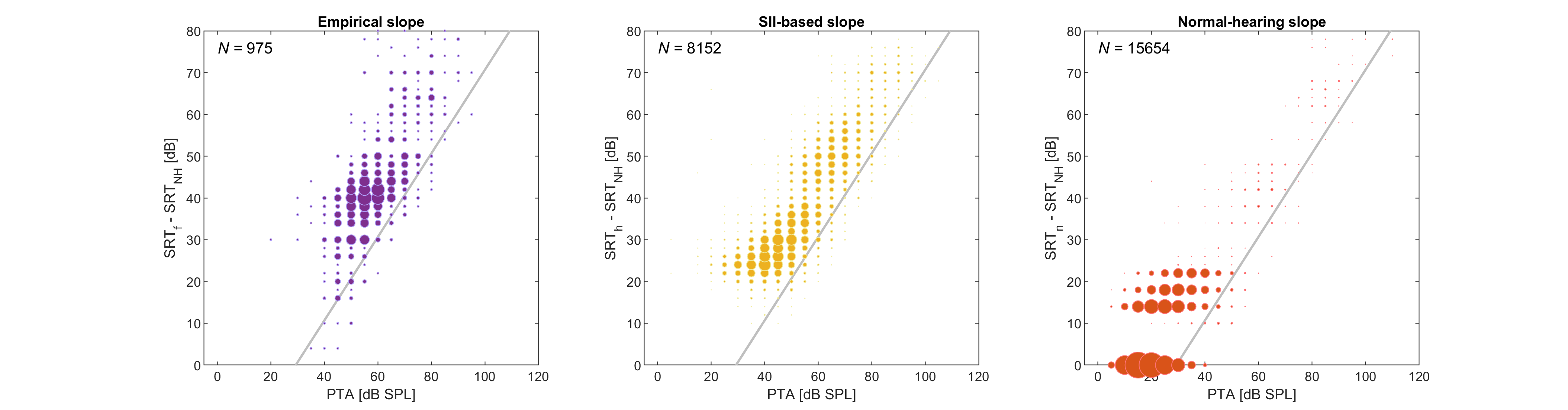}
  \caption{$SRT$ (depicted as $SRT$ loss, the difference to normal-hearing reference $SRT_{NH}$) over $PTA$ for (A) the empirical slope estimation, (B) the SII-slope estimation, and (C) the psychometric function fit with normal-hearing slope and $WRS_{max}=100\%$. The marker size indicates the number of patients in logarithmic scaling. The grey line represents the diagonal where $SRT$ corresponds to $PTA$. } 
\label{fig:srt_pta}
\end{figure}

Figure \ref{fig:srt_pta} shows the estimated SRTs for the three procedures, for the respective possible patient groups as characterized above. The SII-slope-based procedure was also conducted for the fully-determined patients, therefore panel (B) contains all patients that were depicted in yellow and violet in Figure \ref{fig:maxSI_pta}. In comparison, the SRTs obtained with the empirical and the SII-based procedure cover a similar area in the plots, with a larger variation in $SRT$ for a given $PTA$ being obtained with the empirical slope estimation. Most of the patients yield an $SRT$ loss of 40 dB SPL at a $PTA$ of 50-60 dB SPL, while the SII-based procedure shows a maximum between 20 and 30 dB $SRT$ loss at a corresponding $PTA$ of 30 to 50 dB SPL. For the normal-hearing slope procedure, more than half of the patients (56 \%) show a (maximum) $SRT$ loss close to 0 dB. About 43\% show an elevated $SRT$ loss in the range from 15 to 25 dB, while a few patients (with the FMST measured at higher levels) show higher $SRT$ loss, linearly increasing with $PTA$. 
% red: SRT loss = 0: 8710/15654 = 0.56
% between 15 and 25: 6800/15654 = 0.43

The median $SRT$ error was $\Delta SRT_f = 13.68~\text{dB}$ ($IQR = [10.78~18.60]~\text{dB}$) for the $SRT_f$ estimation, and $\Delta SRT_h = 4.61~\text{dB}$ ($IQR = [3.98~5.40]~\text{dB}$) for $SRT_h$. For the $SRT_n$ patients ($N_n = 15506$), the median $SRT$ error was $\Delta SRT_n = 0.71~\text{dB}$ ($IQR = [0.71~14.34]~\text{dB}$). Note that the calculation differed for $SRT_n$ patients, and the error only denotes the maximum $SRT$ deviation towards lower values compared to the estimated $SRT_n$. Therefore, the estimated median shows that more than half of the patients in this group are normal-hearing with $SRT$ loss below 1 dB. 

\subsection{Comparison between empirical and SII-slope-based SRT estimation procedures} % -> Plausibility
\label{sec:comparison}
\textbf{Fully-determined patients}

\begin{figure}[H]
\centering
 \includegraphics[width=1.1\linewidth]{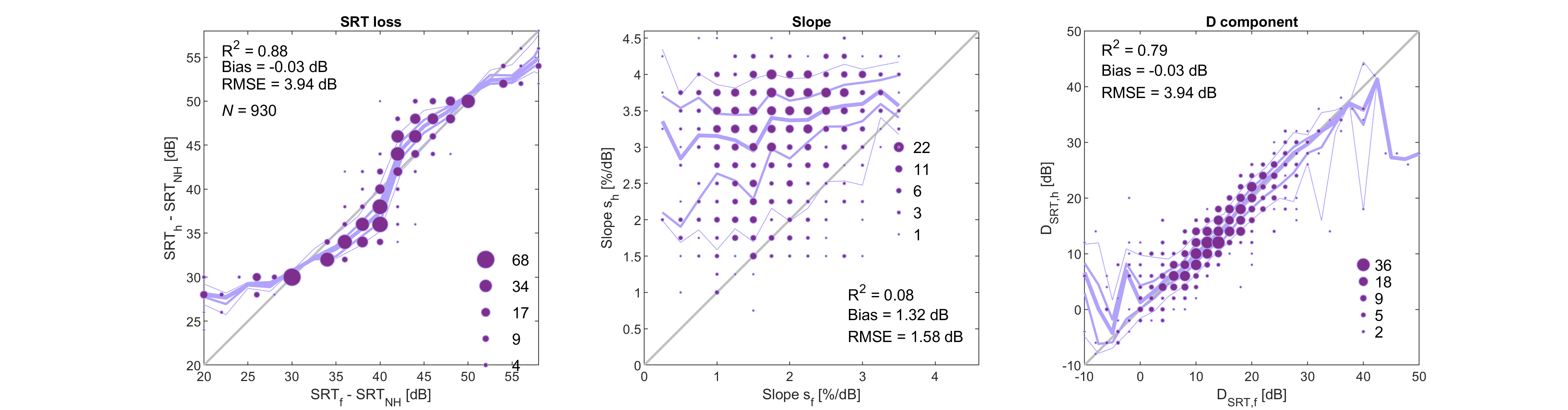}
  \caption{Comparison of empirical slope and SII-slope-based estimation procedure. (A) $SRT$, (B) slope, and (C) $D$ component. The marker size indicates the number of patients in logarithmic scaling. Violet lines represent percentiles of the respective SII-slope-based variable for given empirical variable ranges, the thicker line represents the median.}
\label{fig:srt_2procedures}
\end{figure}

To investigate the applicability and properties of the proposed SII-based procedure, the two procedures were compared for fully-determined patients, for which both procedures were conducted. Figure \ref{fig:srt_2procedures} depicts the $SRT$, slope, and $D$ components obtained with both procedures, for $N=930$ common patients. The comparison of SRTs provides a good correlation between the two procedures, without bias and with $RMSE = 3.94~\text{dB}$. 

In contrast, the corresponding slopes show large deviations, with generally higher SII-based slope and no clear relationship. This divergence merits attention given that $SRT$ and slope are linked in their estimation through linear fitting - accurate $SRT$ estimates typically require correct slope values. Here, the SII-slope-based procedure produced an unexpected outcome: the slope was estimated from the SII slope, therefore the underlying assumptions of the SII influence the slope, and linking it to the only available data point in the \textit{slope area} leads to a good $SRT$ estimation without the possibility for empirical verification of the slope. The $SRT$ estimation is influenced by the measured level and its corresponding $WRS$ value, which serve as anchor point for the linear fit. This is visible in Figure \ref{fig:srt_2procedures} (A), where the variance of the $SRT$ depends on the $SRT$ loss. For presented levels of 60 and 80 dB SPL (corresponding to 31 and 51 dB $SRT$ loss), the SRTs are the same since the SII-based procedure used a data point with $WRS = 50\%$. In between, the variance increases with increasing distance in speech intelligibility, corresponding to the derived error $\Delta SRT_h$ according to Equation \ref{eq:delta_srt}. The increasing difference between the two estimation procedures for $SRT$ loss below the lowest measured level of 60 dB SPL relates to the same concept since the $SRT_h$ estimation depends on $(WRS_i-50\%)$.

The median slope error was $\Delta s_f = 1.57~\text{\%/dB}$ ($IQR = [1.43~1.68]~\text{\%/dB}$) for the $SRT_f$ estimation. The median relative slope error was $\Delta s / s = 0.86$ ($IQR = [0.64~1.17]$), the slope error was nearly as high as the slope estimate itself. The slope error for the $SRT_h$ estimation was $\Delta s_h = 0.0012~\text{\%/dB}$ for all patients (cf. Section \ref{sec:errors}). 
In summary, higher errors are obtained for the empirical procedure as compared to the SII-slope-based procedure, influencing the reliability of the obtained $SRT$ and especially the slope and hence their role as ground truth. 

The $D$ components $D_{SRT,f}$ and $D_{SRT,h}$ show a slightly lower correlation than the $SRT$, but otherwise similar behavior, that is due to their estimation of subtracting the $PTA$ from the $SRT$. The corresponding median errors were $\Delta D_{SRT,f} = 8.68~\text{dB}$ ($IQR = [5.65~13.53]~\text{dB}$) and $\Delta D_{SRT,h} = -0.50~\text{dB}$ ($IQR = [-1.27~0.27]~\text{dB}$), which is 5 dB below the respective errors $\Delta SRT_f$ and $\Delta SRT_h$. 

\textbf{Generalization to half-determined patients}

For half-determined patients, the SII-slope-based procedure could not be compared to the empirical slope procedure. To test if the found relationships between the two $SRT$ estimation procedures can also be expected for the half-determined patients, we compared the distributions of the measured data, namely $PTA$ and $WRS_{max}$, as displayed in Figures \ref{fig:histo_pta}, \ref{fig:histo_maxsi}, and Table \ref{tab:dist_comp} in the appendix. 
The $WRS_{max}$ distributions exhibited a high degree of overlap ($\eta_{WRS_{max}}=0.96$) between groups, with no significant differences in either their means or overall distributions.
The $PTA$ distributions showed significant differences in both mean and distribution, though maintaining substantial overlap ($\eta_{PTA}=0.71$). Fully-determined patients exhibited a narrower $PTA$ range, starting around $40~\text{dB SPL}$ but reaching the same maximum. This pattern reflects the measurement protocol: patients with higher $PTA$ might require more measurement levels to reach $WRS_{max}$. For example, a patient with $PTA = 40~\text{dB SPL}$ might show $WRS < 50\%$ at $60~\text{dB SPL}$ and not reach $WRS_{max}$ at $80~\text{~dB SPL}$, necessitating a third measurement. In contrast, a patient with $PTA = 20~\text{dB SPL}$ likely reaches $WRS > 50\%$ at $60~\text{dB SPL}$. These differences in data availability appear to stem from the measurement protocol of using progressively increasing levels, rather than indicating systematic differences between half- and fully-determined patients. Therefore, we consider the SII-slope based procedure equally applicable to both groups.

\subsection{Prediction of the difference between SII-based ($SRT_h$) and empirical slope SRT ($SRT_f$)}

To assess whether the observed differences between $SRT_f$ and $SRT_h$ could predict $SRT_h$ for all half-determined patients, we built a generalized linear model (GLM), using the slope $s_h$ and the difference in speech intelligibility between the measured $WRS$ and the $SRT$ $(WRS_i-50 \%)$ as predictors. The model, validated using 10-fold cross-validation, yielded significant effects for all predictors ($p < 0.05$, cf. Table \ref{tab:glm_results}). The estimated mean parameters were $\beta_0 = 1.522~\text{dB}$ (intercept), $\beta_1 = -0.515~\text{dB}^2/\%$, and $\beta_2 = 0.146~\text{dB}/\%$. Based on these coefficients, a 10\% change in $WRS$ corresponds to a 1.5 dB change in estimated SRT, while a 0.5 \%/dB change in SII-based slope results in a 0.25 dB $SRT$ change. The cross-validation yielded $RMSE_{CV} = 3.26~\text{dB}$.

\begin{figure}%[H]
\centering
\includegraphics[width=0.5\linewidth]{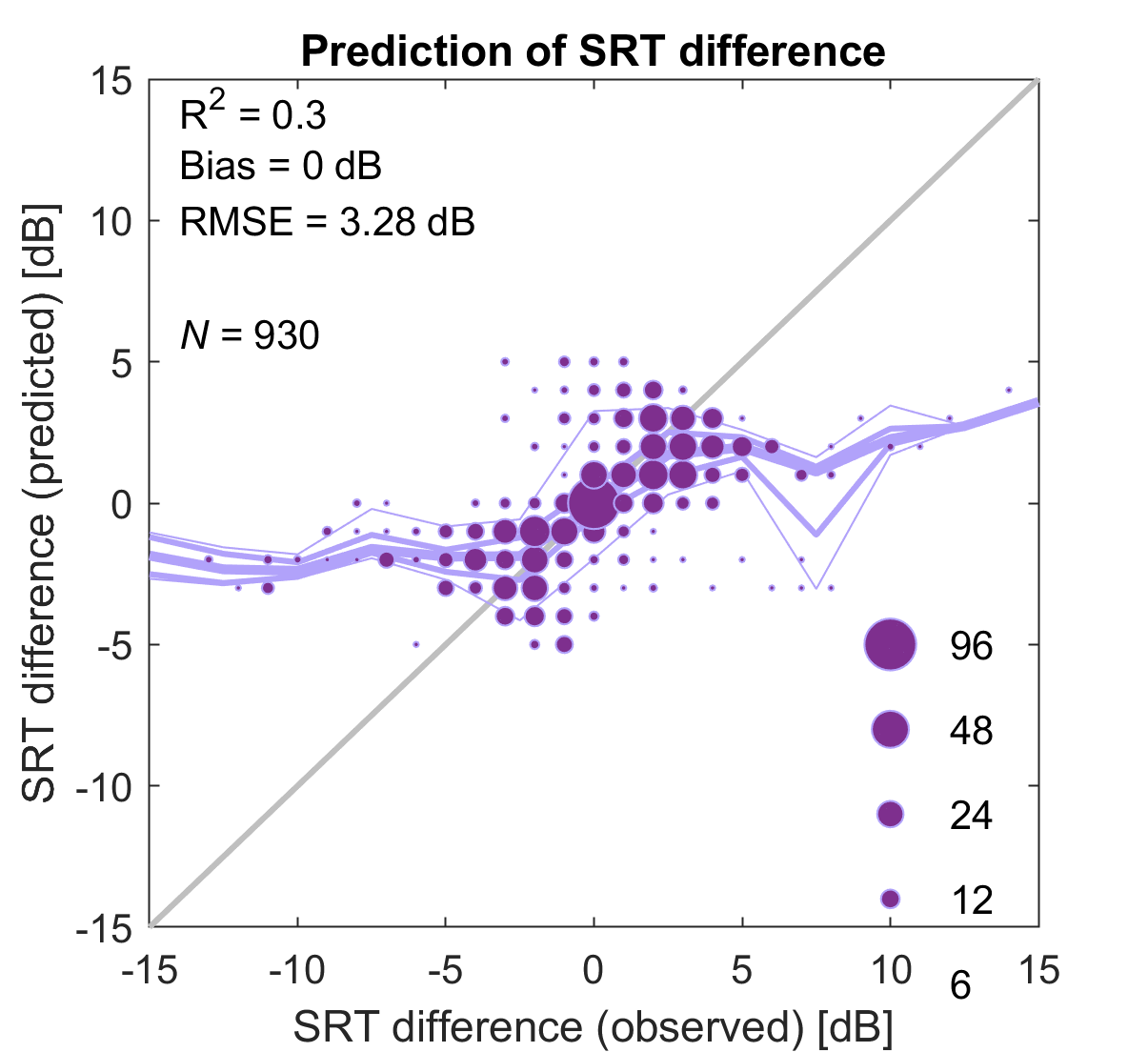}
  \caption{Predicted vs. observed $SRT$ difference ($SRT_f - SRT_h$) for fully-determined patients. The marker size indicates the number of patients in logarithmic scaling. Lighter colored lines represent the $10^{th}$, $30^{th}$, $50^{th}$, $70^{th}$, and $90^{th}$ percentiles of the respective predicted $SRT$ difference for given observed $SRT$ differences, the thickest line represents the median.}
\label{fig:glm1}
\end{figure}

Figure \ref{fig:glm1} shows the predicted compared to the observed $SRT$ differences. All predicted differences are in the range from -5 to 5 dB, larger (and rare) observed differences are not well-represented by the prediction. A poor correlation of $R^2 = 0.3$ was achieved. In the middle area [-5 5] dB, the minimum and maximum predicted $SRT$ differences also correspond to -5 and 5 dB, respectively. Although they seem to scatter a lot, the depicted percentiles show that the largest part of patients is predicted within a range of $\pm 2~\text{dB}$ ($10^{th}$ to $90^{th}$ percentile). In summary, the prediction is feasible and relatively accurate for frequently occurring data combinations in the range of $\Delta SRT =\pm 2.5~\text{dB}$, but not in the full range of $SRT$ differences that were observed. 

\begin{table}[H]
    \centering
               \begin{tabular}{cccccc}
               \toprule
               $k$ & Parameter & Estimate & SE & tStat & $p$  \\ 
                        \midrule
                         \multirow{3}{*}{1} & $\beta_0$ & 1.914 & 0.486 & 3.936 & \textbf{0} \\ 
                          & $\beta_1$ & -0.628 & 0.149 & -4.209 & \textbf{0} \\ 
                          & $\beta_2$ & -0.144 & 0.008 & -19.113 & \textbf{0} \\ 
                         \midrule
                         \multirow{3}{*}{2} & $\beta_0$ & 1.444 & 0.482 & 2.999 & \textbf{0.003} \\ 
                          & $\beta_1$ & -0.505 & 0.148 & -3.404 & \textbf{0.001}\\ 
                          & $\beta_2$ & -0.147 & 0.007 & -19.670 & \textbf{0} \\ 
                         \midrule
                         \multirow{3}{*}{3} & $\beta_0$ & 1.657 & 0.483 & 3.429 & \textbf{0.001} \\ 
                          & $\beta_1$ & -0.553 & 0.149 & -3.715 & \textbf{0} \\ 
                          & $\beta_2$ & -0.144 & 0.008 & -19.189 & \textbf{0} \\ 
                         \midrule
                         \multirow{3}{*}{4} & $\beta_0$ & 1.421 & 0.481 & 2.951 & \textbf{0.003} \\ 
                          & $\beta_1$ & -0.495 & 0.149 & -3.329 & \textbf{0.001} \\ 
                          & $\beta_2$ & -0.146 & 0.008 & -19.136 & \textbf{0} \\ 
                         \midrule
                         \multirow{3}{*}{5} & $\beta_0$ & 1.572 & 0.498 & 3.158 & \textbf{0.002} \\ 
                          & $\beta_1$ & -0.522 & 0.153 & -3.406 & \textbf{0.001} \\ 
                          & $\beta_2$ & -0.152 & 0.008 & -19.703 & \textbf{0} \\ 
                         \midrule
                         \multirow{3}{*}{6} & $\beta_0$ & 1.297 & 0.480 & 2.703 & \textbf{0.007} \\ 
                          & $\beta_1$ & -0.430 & 0.148 & -2.906 & \textbf{0.004} \\ 
                          & $\beta_2$ & -0.144 & 0.008 & -19.090 & \textbf{0} \\ 
                         \midrule
                         \multirow{3}{*}{7} & $\beta_0$ & 1.357 & 0.495 & 2.743 & \textbf{0.006} \\ 
                          & $\beta_1$ & -0.465 & 0.153 & -3.043 & \textbf{0.002} \\ 
                          & $\beta_2$ & -0.148 & 0.008 & -19.107 & \textbf{0} \\ 
                         \midrule
                         \multirow{3}{*}{8} & $\beta_0$ & 1.468 & 0.484 & 3.036 & \textbf{0.002} \\ 
                          & $\beta_1$ & -0.494 & 0.150 & -3.286 & \textbf{0.001} \\ 
                          & $\beta_2$ & -0.145 & 0.008 & -18.874 & \textbf{0} \\ 
                         \midrule
                         \multirow{3}{*}{9} & $\beta_0$ & 1.576 & 0.480 & 3.283 & \textbf{0.001} \\ 
                          & $\beta_1$ & -0.540 & 0.149 & -3.629 & \textbf{0} \\ 
                          & $\beta_2$ & -0.145 & 0.008 & -19.215 & \textbf{0} \\ 
                         \midrule 
                         \multirow{3}{*}{10} & $\beta_0$ & 1.517 & 0.491 & 3.089 & \textbf{0.002} \\ 
                          & $\beta_1$ & -0.515 & 0.152 & -3.393 & \textbf{0.001} \\ 
                          & $\beta_2$ & -0.144 & 0.008 & -18.501 & \textbf{0} \\ 
                         \midrule
                         \multirow{3}{*}{\textbf{mean}} & \textbf{$\beta_0$} & 1.522 & 0.486 & 3.133 & \textbf{0.003} \\ 
                          & \textbf{$\beta_1$} & -0.515 & 0.150 & -3.432 & \textbf{0.001} \\ 
                          & \textbf{$\beta_2$} & -0.146 & 0.008 & -19.160 & \textbf{0} \\
                          \bottomrule \\
                          
                         \end{tabular}
    \caption{GLM results for the prediction of ($SRT_f - SRT_h$), using a normal model distribution. The results for the 10 cross-validation folds and the mean of parameters are reported. $\beta_0$ corresponds to the intercept, $\beta_1$ to the slope, and $\beta_2$ to $(WRS_i-50 \%)$. $p$ values represented in \textbf{bold} are significant at $\alpha = 0.05$ level.}
    \label{tab:glm_results}
\end{table}

\subsection{Supra-threshold hearing loss components based on the two interpretation modes}

\begin{figure}%[H]
\centering
\includegraphics[width=1.1\linewidth]{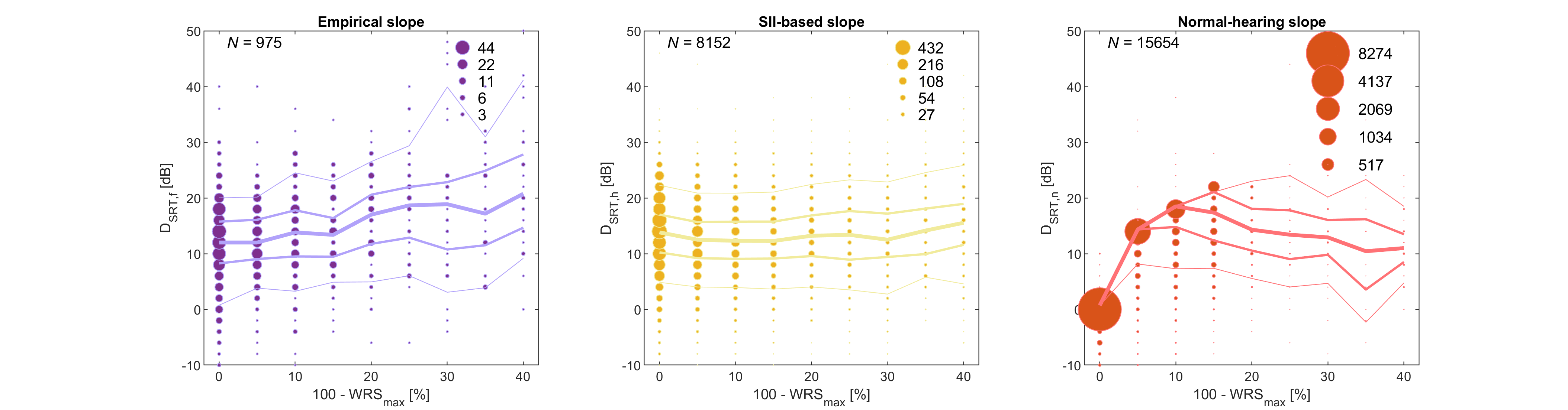}
  \caption{$D_{SRT}$ estimated from different $SRT$ procedures over discrimination loss, for (A) the empirical slope estimation, (B) the SII-slope estimation, and (C) the psychometric function fit with normal-hearing slope and $WRS_{max}=100\%$. The marker size indicates the number of patients in logarithmic scaling. Lighter colored lines represent the $10^{th}$, $30^{th}$, $50^{th}$, $70^{th}$, and $90^{th}$ percentiles of the respective D component for given discrimination loss, the thickest line represents the median.}
\label{fig:D_maxSI}
\end{figure}

Both \textit{interpretation modes} provide a means to characterize a supra-threshold hearing loss component beyond audibility. The discrimination loss ($100\% - WRS_{max}$) is clinically interpreted as the "loss of information-carrying capacity" \citep{halpin_article_2009}, a $D$ component according to \cite{plomp_auditory_1978} can be estimated from $SRT$ and $PTA$. To investigate if a relationship between these two supra-threshold interpretations exists in the measured data, the two quantities were compared. Figure \ref{fig:D_maxSI} shows the results for the three $SRT$ estimation procedures. 

For the empirical slope procedure, the median $D$ component $D_{SRT,f}$ slightly increases with increasing discrimination loss. For each discrimination loss value, the variance of obtained $D$ components is very large. Although the change in the median from 0 to 40\% discrimination loss is about 8 dB, this is still in the order of magnitude of the error $\Delta D_{SRT,f}$, and the number of patients is very low for high discrimination loss, especially compared to the large number for 0\% discrimination loss. The $10^{th}$ and $90^{th}$ percentiles span a larger range of 30 dB at 0\% discrimination loss compared to a range of about 20 dB with increasing discrimination loss. 

For the SII-slope-based procedure, in general a similar picture is obtained, based on a higher number of patients. In contrast, the median $D$ component $D_{SRT,h}$ is constant and therefore not depending on the discrimination loss. If there is a relationship in the empirical slope procedure (hidden by the error), this relationship is not captured by the $D$ components from the SII-slope-based procedure.

For the normal-hearing slope procedure, the $D$ component $D_{SRT,n}$ indicates, as for the corresponding $SRT_n$, the maximum possible $D$ component for the respective patient. $D_{SRT,n}$ equals 0 for $0\%$ discrimination loss, which corresponds to the normal-hearing psychometric function and SRT. With increasing discrimination loss, the maximum $D_{SRT,n}$ increases, with the largest part of patients showing the respective maximum and therefore an $A$ component of $A=0~\text{dB}$. Fewer patients show lower maximum $D$ components, which corresponds to patients with a $PTA$ larger than $SRT_{NH}$, hence a higher $A$ component.
In summary, the given data and results do not allow to find a relationship between $D$ component and discrimination loss that seems to provide the same interpretation of the two \textit{interpretation modes}.

%%%%%%%%%%%%%%%%%%%%%%%%%%%
%%%%%%%%%%%%%%%%%%%%%%%%%%%
%%%%%%%%%%%%%%%%%%%%%%%%%%%

%%%%%%%%%%%%%%%%%%%%%%%%%%%
%%%%%%%%%%%%%%%%%%%%%%%%%%%
%%%%%%%%%%%%%%%%%%%%%%%%%%%
\section{Discussion}
%%%%%%%%%%%%%%%%%%%%%%%%%%%
%%%%%%%%%%%%%%%%%%%%%%%%%%%
%%%%%%%%%%%%%%%%%%%%%%%%%%%
The main aim of this paper was to characterize the relationship between the two \textit{interpretation modes} of speech tests, $SRT$ and $WRS_{max}$, based on an existing clinical database. To this end, we investigated for which patients (and corresponding data availability in a typical clinical database) SRTs can be estimated from data targeting the maximum discrimination ability. The two \textit{interpretation modes} $SRT$ and $WRS_{max}$ are related to different parts of the underlying psychometric function and therefore to different possible statements and interpretations. Assessing the relationship between the two interpretation modes is important, e.g., if knowledge from databases is to be compared that contain measurement outcomes in different interpretation modes. 

\subsection{SRT estimation}

We evaluated whether reliable $SRT$ values could be determined from an existing clinical database that targeted maximum discrimination ability. In general, the feasibility and selection of possible $SRT$ estimation method depends on the available data points (level and $WRS$). Our analysis revealed that $SRT$ estimation based on measured data was possible for only 3.4\% of patients (\textit{fully-determined}). The proposed model-driven procedure expanded this capability to one quarter of patients (\textit{half-determined}), while for the remaining patients only an estimation of maximum $SRT$ based on a normal-hearing slope assumption was possible (\textit{undetermined}).

The $SRT$ estimation for fully-determined patients provides the optimal approach and could be considered as ground truth, since deriving an $SRT$ from a slope corresponds to the characterization of the linear part of the underlying psychometric function, capturing the relationship between the two parameters. However, the median error of $\Delta SRT_f = 13.68~\text{dB}$, which is originating from the high $WRS$ confidence intervals of the FMST \citep{holube_modelling_nodate}, needs to be taken into account for interpretation and influences reliability (\textbf{RQ 1}).

The SII-slope-based procedure was proposed for half-determined patients. To assess its applicability, it was compared to the empirical slope procedure for fully-determined patients, yielding a good correlation of $SRT$ and corresponding $D$ components between the two procedures. In contrast, the slope deviated largely between procedures, showing that the SII-based slope cannot appropriately represent individual empirical slopes (\textbf{RQ 2}). On the other hand, the median relative slope error of the empirical slope procedure ($\Delta s/s = 0.86$) was very high and does not provide a good slope estimate, which should have a relative error below 0.25 \citep{brand2002efficient}. The slope of the SII-level curve represents the increase of audible cues that are available above the hearing threshold with increasing speech level \citep{ansi_1997}. Therefore, the slope is mainly associated with the individual audiogram, with different frequencies being weighted by the band importance function used in the SII calculations. With increasing level, the speech signal becomes more and more audible, with a steeper slope for flat audiogram shapes and a shallower slope if the hearing threshold differs across frequencies, due to the cumulative behavior of the SII curve. The SII may overestimate speech intelligibility and the corresponding slope since it does not incorporate supra-threshold deficits that additionally affect speech intelligibility \citep{hulsmeier_inference_2021}, which fits to the generally higher observed SII-based slope compared to the empirical slope (cf. Figure \ref{fig:srt_2procedures}).  

Despite the inaccuracies in the slope, the $SRT$ was comparable between the two procedures. This seems to be related to the data collection itself, that is, the choice of levels and the fact that the $SRT$ needs to lie between the level corresponding to $WRS_{max}$ and the lowest presented level chosen based on audibility (monotonically increasing). As visible in the error estimation (Equation \ref{eq:delta_srt}), the error depends on ($WRS_i - 50\%$) and therefore on the obtained speech intelligibility at the (pre-defined) presented level. Although a more accurate slope estimate would increase the accuracy of $SRT$ estimation, the SII-based slopes allow for a good $SRT$ estimate. This is also supported by the prediction of the $SRT$ difference between the two procedures, where the predictor variables ($WRS_i - 50\%$) and slope $s_h$ were found to be significant for the prediction, and the magnitude of fitted model parameters showed that ($WRS_i - 50\%$) has a much larger influence on the $SRT$ difference than a wrong slope estimate (\textbf{RQ 3}). The above considerations apply both for fully-determined as well as half-determined patients due to their equal dependence on the SII and the data collection, which was also supported by the comparison of $PTA$ and $WRS_{max}$ distributions between the two groups. In summary, the SII-based procedure can be employed to obtain a plausible $SRT$, considering its error and properties.   

\subsection{Supra-threshold interpretation}
\label{sec:disc_supra}
The supra-threshold hearing deficits that are assumed for the two \textit{interpretation modes} did not show a consistent picture in the present analysis. Here, it was not clear from literature if exactly the same interpretation of supra-threshold deficits was to be expected: While the $WRS_{max}$ is typically described as "loss in information-carrying capacity" \citep{halpin_article_2009} and is clinically used to estimate if a benefit from a hearing device can be expected, the $D$ component according to Plomp \citep{plomp_auditory_1978} represents all deficits that are not covered by audibility, thus related to the pure-tone audiogram (or more precisely the $PTA$). Hence, the $D$ component may involve different deficits, for example related to distortions affecting speech intelligibility or to loudness perception \citep{sanchez_lopez_data-driven_2018, saak_flexible_2022}. 
With increasing discrimination loss calculated from $WRS_{max}$, the (median) $D$ components as derived from the estimated SRTs were constant, that is, not or only slightly (for the empirical slope $SRT$ estimation procedure) depending on the discrimination loss. At the same time, the variance of the $D$ component was very large, with the especially interesting finding that this large range of $D$ components was also found for a discrimination loss of 0\%. This means that a supra-threshold deficit is captured in the $D$ component, which is not captured in the discrimination loss. Therefore, the outcome variables $SRT$ and $(100\% - WRS_{max})$ either represent different types of supra-threshold deficits, or the difference is due to the levels at which the interpretation modes are "operating", namely more relevant for daily life communication, or close to the uncomfortable level of individual participants. 
In addition, the method of estimating Plomp's $D$ component could influence the comparison of interpretation modes: given that the current estimation method was based on the pure-tone average, frequency-specific audibility was not considered. Calculating a frequency-specific $D$ component would be one way to rule out the influence of frequency-specific audibility and to assess the supra-threshold deficits in more detail.
However, the comparison of supra-threshold components was subject to high errors in both interpretation modes, originating from the test-retest reliability of the FMST. The tendency of a slightly increasing $D$ component with increasing discrimination loss for the empirical $SRT$ estimation procedure could give the idea that the supra-threshold $D$ component is better reflected in the real slope than in the SII-based slope, but this cannot be verified due to the error of the speech test (\textbf{RQ 4}).  

\subsection{Implications for speech intelligibility measurements in clinical practice} 

Some recommendations for speech intelligibility measurements in clinical practice can be derived from the outcomes of the present analysis.  
Characterizing not only the maximum discrimination score, but also the $SRT$ for a patient provides complementary information about supra-threshold hearing deficits.
Despite the shown feasibility of the SII-slope-based $SRT$ estimation, the real individual slope (and corresponding $SRT$) can only be known if two FMST test lists are conducted that lead to word recognition scores in the linear part of the underlying psychometric function. 
To achieve this, we recommend measuring three test lists in total for every patient. 
The measured levels should be chosen appropriately to allow for $SRT$ and slope estimation based on two data points in the assumed linear range of the psychometric function, as well as to characterize the maximum word recognition score. While the level choice for different test lists was assessed retrospectively in this study, this knowledge could also be used to develop a new adaptive procedure to suggest optimized levels during the measurement.  
However, the large errors of the FMST lead to large errors of the estimated $SRT$. The dependence of the confidence intervals on the obtained $WRS$ \citep{holube_modelling_nodate} results in a larger $\Delta WRS$ at data points used for $SRT$ estimation compared to the data points at which $WRS_{max}$ is obtained. As an additional source of error, the speech intelligibility obtained at the same level varies across test lists \citep{baljic2016evaluation}, which complicates comparability across test lists and even more the usage of the single FMST words in an adaptive procedure. Therefore, a further (long-term) recommendation would be to choose a more accurate speech test for patient characterization. A candidate could be the Matrix Sentence Test, which was proposed by \cite{kollmeier2011horgerateindikation} to be used for the assessment of hearing device benefit, and which could be measured at a noise level of 45 dB allowing for an effective measurement in quiet or in noise depending on the hearing threshold of the respective patient. 

The two \textit{interpretation modes} have different relevance for different patient groups as it is neither important nor possible to estimate an $SRT$ for all patients. 
These patient groups correspond to those with different data availability identified in the present clinical database.  
For patients with low maximum word recognition score, their low performance was the reason why no $SRT$ could be estimated (blue group in Figures \ref{fig:data_flowchart} and \ref{fig:maxSI_pta}). These patients mainly correspond to CI candidates which should be further characterized according to the guidelines \citep{ci_guidelines}, and $WRS_{max}$ is sufficient to state the severity of their hearing loss, which includes a large audibility component. At the other extreme, the largest part of patients with $WRS_{max} \geq 80\%$ at 60 dB SPL was only characterized by one test list (\textit{undetermined slope} patients), since the high performance is already sufficient for not being eligible for any hearing aid according to the guidelines \citep{ha_guidelines}. However, as discussed in Section \ref{sec:disc_supra}, even for these patients a supra-threshold $D$ component might be present, therefore a more detailed characterization with either more test lists of the FMST or a more precise test could be important to assess hearing deficits beyond audibility.  
For the \textit{fully-} and \textit{half-determined slope} patients, the characterization of the $SRT$ should provide a benefit as discussed above.          
\subsection{Research on clinical databases}

Analyzing existing clinical databases offers distinct advantages over traditional research studies. While typical research studies collect data for specific purposes with well-defined participant groups and a potentially larger number of different tests, analysis on clinical databases allows for the assessment of clinical practice and patterns contained in the data, including standard tests as established for real patient characterization. Large patient numbers enable robust statistical analysis of relationships, potentially offering a representative picture of the hearing-impaired population when multiple databases are combined in the future.  
However, retrospective analysis of clinical data presents unique challenges. Ensuring data quality becomes more complex when dealing with routine clinical data, which may be incomplete or have patient information split across multiple files. Depending on the planned analysis, data from different computers or measurement devices need to be integrated, and data protection and privacy have to be ensured \cite{cantuaria_hearing_nodate}. 

In our study, we addressed these challenges through several measures. The well-structured research database of Hanover Medical School, developed, maintained, and refined over 20 years, has undergone extensive data cleaning. The inclusion criteria for the present analysis provide an additional filter for patient data that are to be included. Here, our inclusion criteria were not overly restrictive, since a methodological question about relationships within patients was investigated, and data consistency was ensured by requiring same-day measurements for the different tests. We employed a remote analysis approach that preserved patient privacy by keeping data on the server of Hanover Medical School, and by collaboratively developing the analysis scripts on synthetic example data. 

\subsection{Limitations and outlook}
While using a clinical database provided valuable insights, a direct comparison of $SRT$ estimation procedures was limited to 3.4\% of the included patients (fully-determined), which however corresponds to the high absolute number of $N=930$ patients. Validation with data sets containing complete psychometric functions, such as those described by \cite{hoppe_age-related_2022}, would strengthen our findings. 

Our choice of the SII model, while suitable for deriving slope estimates from level-dependent SII curves and individual audiogram data, has inherent limitations. The SII doesn't account for supra-threshold deficits beyond audibility. Alternative speech intelligibility models such as the Framework for Auditory Discrimination Experiments (FADE; \cite{marc_rene_schadler_matrix_2015}) might better predict Plomp's $D$ component, but require additional individualization parameters, for example from a tone-in-noise measurement \citep{hulsmeier_how_2022}, which are not available in our clinical database. The SII was chosen primarily to incorporate a consistent relationship between audiogram and speech intelligibility. 

The high uncertainty of the FMST measurements \citep{holube_modelling_nodate} constrained our ability to compare supra-threshold deficits between $SRT$ and $WRS_{max}$ interpretations. This limitation suggests the need for future studies using more precise speech tests to better understand these relationships. 

Future work can build on the findings of this study, which provides important learnings about the two interpretation modes, the potential to estimate an $SRT$ from data targeting at $WRS_{max}$, and its dependencies on the error of the underlying speech test. Towards a common interpretation of speech tests, the SII-slope-based procedure can be employed on data in the $WRS_{max}$ interpretation mode to estimate an $SRT$ and the respective corresponding error. Based on this, other differences between speech tests can be compared and investigated. 

%%%%%%%%%%%%%%%%%%%%%%%%%%%
%%%%%%%%%%%%%%%%%%%%%%%%%%%
%%%%%%%%%%%%%%%%%%%%%%%%%%%
\section{Conclusions} 

This study investigated the potential for $SRT$ estimation based on clinical data characterizing maximum speech intelligibility. The main conclusions were: 

\begin{itemize}
    \item The clinical data collection allowed for an $SRT$ estimation based on measured data only for a small part of the database (3.4\%), with a median error of 13.68 dB (\textbf{RQ 1}). For one quarter of the included patients, our proposed SII-slope-based $SRT$ estimation was feasible, exhibiting a median error of 4.61 dB. For the remaining patients, only one data point was available, allowing for an estimation of the maximum limit of the SRT. No $SRT$ estimation is feasible for patients with too low maximum word recognition score (\textbf{RQ 2}). 
    \item The SII-slope-based procedure offers a novel, model-based method for estimating $SRT$ when limited data are available. While the $SRT$ was highly correlated to the empirical SRTs, differences in slope were found that can be explained by properties of the SII. The successful $SRT$ estimation can be explained by the parameters used in this procedure, with a higher influence of the obtained word recognition score at the available data point compared to the slope used for the linear fit (\textbf{RQ 3}).
    \item A potential relationship between supra-threshold hearing loss components estimated by the two interpretation modes could not be identified due to the low test-retest reliability of the FMST. The observed large variations can either be related to the error, or, if they were confirmed based on more precise measurements, could mean that (partly) different supra-threshold interpretations can be obtained from the two interpretation modes (\textbf{RQ 4}).  
    \item Our approach demonstrates successful knowledge extraction from a clinical database while maintaining awareness of inherent limitations. Based on the gained insights, we recommend routinely measuring three test lists at appropriate levels, if time allows. However, a more accurate $SRT$ estimation would be facilitated based on speech tests with better test-retest reliability.
    \item In future work, the proposed SII-slope-based $SRT$ estimation procedure can be used in additional studies about comparability and common interpretation of speech intelligibility tests. For this purpose, it is important to compare data in the same interpretation mode and to consider the corresponding errors. 
\end{itemize}

%%%%%%%%%%%%%%%%%%%%%%%%%%%
%%%%%%%%%%%%%%%%%%%%%%%%%%%
%%%%%%%%%%%%%%%%%%%%%%%%%%%
\section{Acknowledgments}
MB was funded by the Deutsche Forschungsgemeinschaft (DFG, German Research Foundation) - Project ID 496819293, and the “Fondation Pour l’Audition” (FPA IDA10). MC and PA were funded by the “Fondation Pour l’Audition” (FPA IDA10). EK and LSM were supported by the Deutsche Forschungsgemeinschaft (DFG, German Research Foundation) under Germany’s Excellence Strategy - EXC 2177/1 - Project ID 390895286. 

\section{Declaration of interest statement}
MB received a payment for an invited talk by Sonova Audiological Care France. No other potential conflict of interest was reported by the authors. 

\section{Data availability}
The research database of Hanover Medical School can be accessed by members and collaborators of the Cluster of Excellence "Hearing4all", via a remote analysis procedure.  
The code for the presented analysis can be found in Zenodo, \url{https://doi.org/10.5281/zenodo.14634515}. 

\section{Author Contributions}
MB: Conceptualization, funding, data curation, formal analysis, investigation, methodology, software, writing--original draft preparation, writing--review and editing.\\
EK: Data curation, software, writing--review and editing.\\
LSM: Investigation, writing--review and editing.\\
PA: Supervision, funding, writing--review and editing.\\
MC: Formal analysis, investigation, writing--original draft preparation, writing--review and editing.\\
All authors have read and agreed to the final version of the manuscript.

%title page; abstract; keywords; main text introduction, materials and methods, results, discussion; acknowledgments; declaration of interest statement; references; appendices (as appropriate); table(s) with caption(s) (on individual pages); figures; figure captions (as a list)

\bibliography{Buhl_srt_bib} 
\bibliographystyle{apalike} % unsrt
%%%%%%%%%%%%%%%%%%%%%%%%%%%
\section{Appendix}
 \label{appendix:app1}

\subsection{Histograms for measured data (corresponding to Section \ref{sec:comparison})}

 \begin{figure}[H]
\centering
 \includegraphics[width=0.5\linewidth]{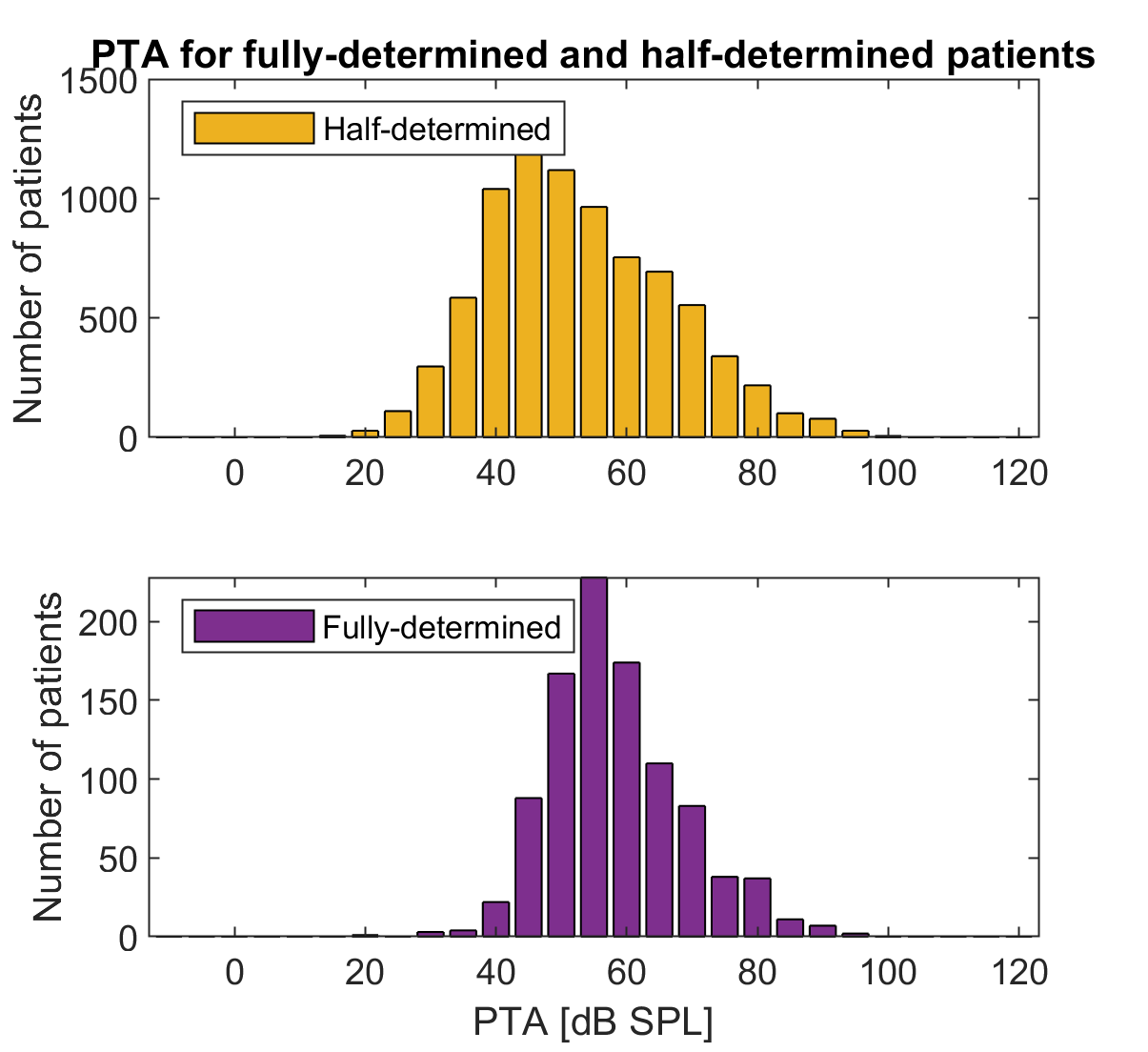}
  \caption{Histograms for the $PTA$ calculated on measured data, for \textit{fully-determined slope} and \textit{half-determined slope} patients.}
\label{fig:histo_pta}
\end{figure}

\begin{figure}[H]
\centering
 \includegraphics[width=0.5\linewidth]{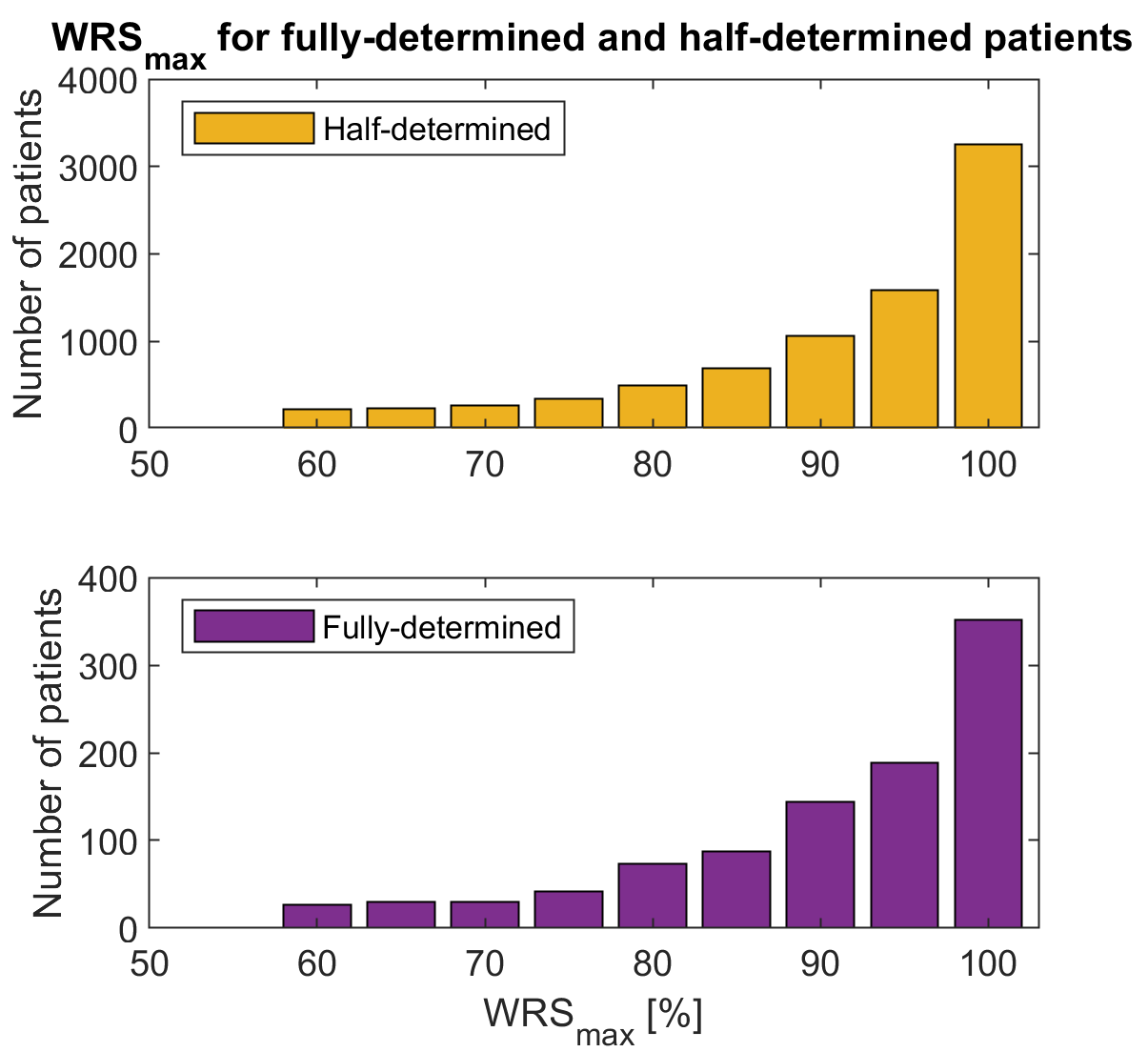}
  \caption{Histograms for the measured $WRS_{max}$ data, for \textit{fully-determined slope} and \textit{half-determined slope} patients.}
\label{fig:histo_maxsi}
\end{figure}

\begin{table}
    \centering
    \begin{tabular}{ccccc}
        \toprule
         &  Overlapping Index &  Welch Test &  Kolmogorov-Smirnov & \\
         \midrule
                    & $\eta$ & $p_{mean}$ & $p_{dist}$ & \\
                    \midrule
        $PTA$       &  $0.71$  & $< 0.001$ & $< 0.001$ \\ %& \textbf{$5.061 \cdot 10^{-34}$} & \textbf{$2.443 \cdot 10^{-63}$} & \\
        $WRS_{max}$ &  0.96  & 0.102 & 0.139 & \\
        \bottomrule \\
    \end{tabular}
    \caption{Statistical test results for the comparison of measured $PTA$ and $WRS_{max}$ distributions between \textit{fully-determined slope} and \textit{half-determined slope} patients.}
    \label{tab:dist_comp}
\end{table}

\end{document}